\documentclass[preprint]{aastex}

\newcommand{\nhat}{{\hat n}}
\newcommand{\COBE}{\textsl{COBE}}
\newcommand{\WMAP}{\textsl{WMAP}}

\newcommand{\refeqnp}[1]{(eq.~[\ref{#1}])}
\newcommand{\refeqnt}[1]{{equation~(\ref{#1})}}
\newcommand{\reffigp}[1]{(Fig.~\ref{#1})}
\newcommand{\reffigt}[1]{{Figure~\ref{#1}}}
\newcommand{\reftbl}[1]{{Table~\ref{#1}}}

\newcommand{\minchisq}{12.5}
\newcommand{\nullchisq}{17.2}
\newcommand{\bestomegal}{0.68}
\newcommand{\deltachisq}{4.7}

\begin{document}

\slugcomment{Submitted to The Astrophysical Journal}

\title{First Year \textsl{Wilkinson Microwave Anisotropy Probe}
(\WMAP\altaffilmark{1}) Observations:\\
Dark Energy Induced Correlation with Radio Sources}

\author{
M. R. Nolta\altaffilmark{2},
E. L. Wright\altaffilmark{3},
L. Page\altaffilmark{2},
C. L. Bennett\altaffilmark{4},
M. Halpern\altaffilmark{5},
G. Hinshaw\altaffilmark{4},
N. Jarosik\altaffilmark{2}, 
A. Kogut\altaffilmark{4}, 
M. Limon\altaffilmark{4,9}, 
S. S. Meyer\altaffilmark{6},
D. N. Spergel\altaffilmark{7},
G. S. Tucker\altaffilmark{8,9},
E. Wollack\altaffilmark{4}
}

\altaffiltext{1}{{\WMAP} is the result of a partnership between Princeton
University and the {\sl NASA} Goddard Space Flight Center. Scientific guidance
is provided by the {\WMAP} Science Team.}
\altaffiltext{2}{{Dept. of Physics, Jadwin Hall, Princeton, NJ 08544}}
\altaffiltext{3}{{UCLA Astronomy, PO Box 951562, Los Angeles, CA 90095-1562}}
\altaffiltext{4}{{Code 685, Goddard Space Flight Center, Greenbelt, MD 20771}}
\altaffiltext{5}{{Dept. of Physics and Astronomy, University of British Columbia, Vancouver, BC  Canada V6T 1Z1}}
\altaffiltext{6}{{Depts. of Astrophysics and Physics, EFI and CfCP, University of Chicago, Chicago, IL 60637}} 
\altaffiltext{7}{{Dept of Astrophysical Sciences, Princeton University, Princeton, NJ 08544}}
\altaffiltext{8}{{Dept. of Physics, Brown University, Providence, RI 02912}}
\altaffiltext{9}{{National Research Council (NRC) Fellow}}

\email{mrnolta@princeton.edu}

\begin{abstract}
The first-year {\WMAP} data, in combination with any one of a number of
other cosmic probes, show that we live in a flat $\Lambda$-dominated
CDM universe with $\Omega_m\approx0.27$ and $\Omega_\Lambda\approx0.73$.
In this model the late-time action of the dark energy, through the integrated
Sachs-Wolfe effect, should produce CMB
anisotropies correlated with matter density fluctuations at $z\la2$
\citep{crittenden/turok:1996}.
The measurement of such a signal is an important independent
check of the model.
We cross-correlate the NRAO VLA Sky Survey radio source catalog
\citep{condon/etal:1998} with the {\WMAP} data in search
of this signal, and see indications of the expected correlation.
Assuming a flat $\Lambda$CDM cosmology, we find $\Omega_\Lambda>0$
(95\% CL, statistical errors only)
with the peak of the likelihood at $\Omega_\Lambda=\bestomegal$, consistent
with the preferred {\WMAP} value.
A closed model with $\Omega_m=1.28$, $h=0.33$, and no dark energy
component ($\Omega_\Lambda=0$),
marginally consistent with the {\WMAP} CMB TT angular
power spectrum, would produce an anti-correlation between the
matter distribution and the CMB.
Our analysis of the cross-correlation of the {\WMAP} data with the
NVSS catalog rejects this cosmology at the $3\sigma$ level.
\end{abstract}

\keywords{cosmic microwave background, cosmology: observations}

\section{Introduction}

The recent {\WMAP} results \citep{bennett/etal:2003}
place on a firm foundation the emerging standard model of cosmology:
a flat adiabatic $\Lambda$-dominated CDM universe.
However, the deficit of power in the CMB anisotropy spectrum on large angular
scales \citep{bennett/etal:2003b,hinshaw/etal:2003,spergel/etal:2003}
is surprising given that $\Lambda$CDM models predict an enhancement 
at $\ell\la10$ due to the late-time integrated Sachs-Wolfe (ISW) effect
\citep{sachs/wolfe:1967}.
The likelihood of observing so little power due to sample variance
is only $0.15\%$ \citep{spergel/etal:2003}.
Thus new physics may be indicated since the nature of the dark energy
is poorly understood.
Cross-correlating the CMB with radio sources provides a direct test for the 
recent acceleration predicted by the favored
$\Lambda$CDM model and observed by the Type 1a supernovae experiments
\citep{perlmutter/etal:1999,riess/etal:1998}.
Thus, this test is an important check of the standard model.

\begin{figure}
\plotone{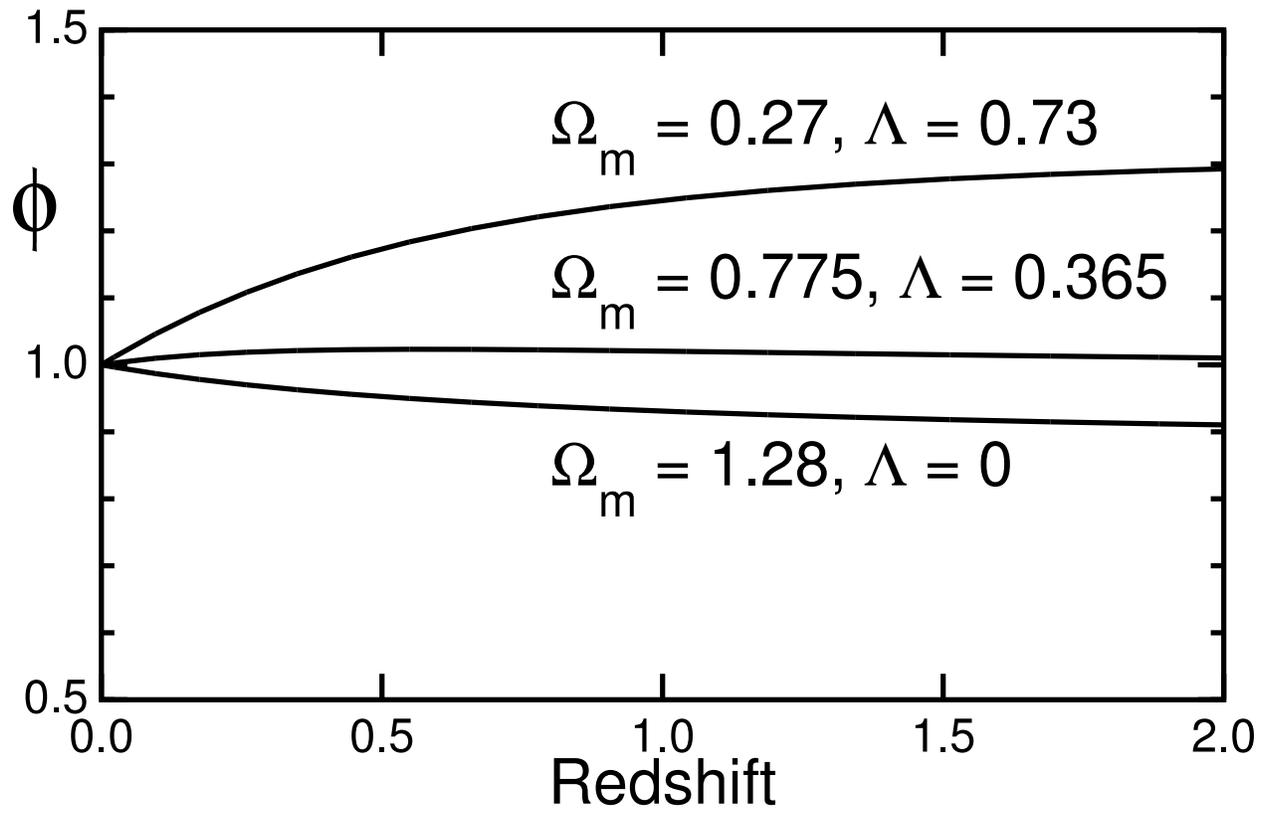}
\caption{\label{fig:phi}
The gravitational potential $\Phi$ as a function of redshift $z$ for
a variety of cosmological models. The models are normalized to unity at
$z=0$.
}
\end{figure}

Any recent acceleration of the universe will cause local gravitational
potentials to decay.
This decay is then imprinted on the CMB as the photons, which were blue-shifted
on infall, suffer less of a red-shift as they climb out of the potential well.
This produces temperature perturbations
\begin{equation}
\frac{\delta T(\nhat)}{T_0}
	= -2\int_0^{\eta_{\rm dec}}{d\eta\,\frac{d\Phi}{d\eta}(\eta\nhat)}
\label{eqn:dT-ISW}
\end{equation}
where $\Phi$ is the Newtonian gravitational potential, $\eta$
is conformal lookback time, and the integral runs from today
($\eta=0$) to the CMB decoupling surface at $z_{\rm dec}=1089$.
\reffigt{fig:phi} shows the recent evolution of $\Phi$ for a variety of
cosmological models.
Since $\Phi$ is related to the matter distribution via the Poisson equation,
tracers of the mass will be correlated with the CMB through the late-ISW effect.
In this paper we correlate the NRAO VLA Sky Survey (NVSS) source catalog
\citep{condon/etal:1998} with the {\WMAP} CMB map in search of the
late-ISW effect.
\citet{boughn/crittenden:2002} performed a similar analysis using
the \COBE/DMR map, and found no correlation.
However, a recent re-analysis by the same authors using the {\WMAP} data
did see evidence for a correlation between the CMB, NVSS,
and the hard X-ray background observed by the HEAO-1 satellite
\citep{boughn/crittenden:2003}.
Here we focus on the implications of this correlation for
dark energy, specifically $\Omega_\Lambda$.

We take as our fiducial model the best fit power-law $\Lambda$CDM model
to the combined \WMAP, CBI, ACBAR, 2dFGRS, and Ly$\alpha$ data sets, with
values
$\omega_m=0.133$, $\omega_b=0.0226$, $n_s=0.96$, $h=0.72$, $A=0.75$, 
and $\tau=0.117$
\citep[Table 7]{spergel/etal:2003}.
Cosmological parameters are drawn from this set unless otherwise noted.

In \S2 we briefly describe the NVSS source catalog and its
auto-correlation function.
In \S3 we describe the cross-correlation between the {\WMAP} CMB map
and the NVSS source map, and relate it to the late-ISW effect.
We conclude in \S4 and discuss effects which could mimic the observed signal.

\section{The NVSS Source Catalog}

The NRAO VLA Sky Survey (NVSS) is a 1.4 GHz continuum survey,
covering the 82\% of the sky with $\delta>-40\arcdeg$ \citep{condon/etal:1998}.
The source catalog contains over $1.8\times 10^6$ sources,
and is 50\% complete at 2.5~mJy. 
Nearly all the sources away from the Galactic plane ($|b|>2\arcdeg$) are
extragalactic.
The bright sources are predominantly AGN-powered radio galaxies and
quasars, while at weaker fluxes the sources are increasingly
nearby star-forming galaxies.

Galaxy counts are a potentially biased tracer of the underlying matter
distribution, and thus the projected number density
of NVSS sources per steradian, $n(z,\nhat)$,
is related to the matter distribution $\delta(z,\nhat)\equiv\delta\rho/\rho$
via
\begin{equation}
n(z,\nhat) \,dz\,d\Omega
= \frac{dN}{dz}\left(1 + b_r(z)\delta(z,\nhat)\right) \,dz\,d\Omega
\end{equation}
where $dN/dz$ is the mean number of sources per steradian at a redshift $z$
and $b_r(z)$ is the radio galaxy bias parameter.
Thus the observed fluctuation on the sky in projected source counts is given by
\begin{equation}
\delta N(\nhat) = \int{dz\, b_r(z)\frac{dN}{dz}\delta(z,\nhat)}.
\label{eqn:dN-NVSS}
\end{equation}
Since we are only interested in clustering on large scales, and hence the
linear regime, the evolution of $\delta$ factors as
$\delta(k,z)=D(z)\delta(k)$, where $\delta(k,z)$ is the Fourier
transform of the matter distribution,
$\delta(k)\equiv\delta(k,0)$ its current value,
and $D(z)$ is the linear growth factor \citep[5.111]{peebles:POPC}.
While generally a function of time, we take the bias to be constant,
as the determination of $dN/dz$ is uncertain 
and we only consider a modest redshift range ($0<z<2$).

\begin{figure}
\plotone{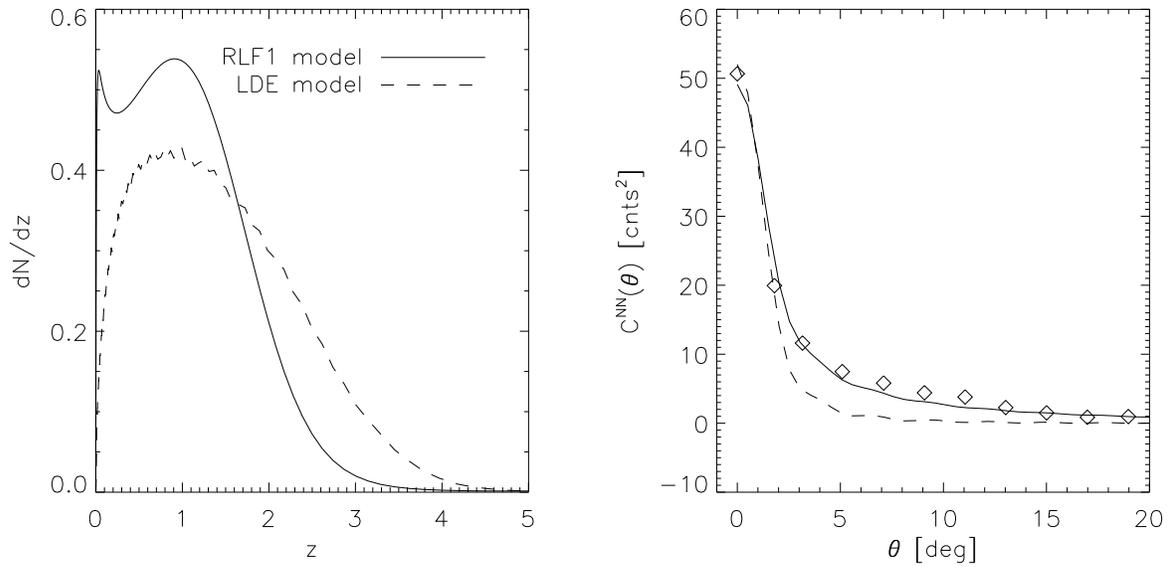}
\caption{\label{fig:dndz}
The adopted $dN/dz$ model (RLF1) for the distribution of NVSS sources
from \citet[DP90]{dunlop/peacock:1990}, normalized to integrate to unity
(left panel).
The small blip at $z\approx0.05$ is spurious, due to breakdown in the
DP90 fitting function.
Also plotted is the ``luminosity/density evolution'' (LDE) model also from
DP90, which is a poor fit to the observed auto-correlation function
(right panel).
}
\end{figure}

While the individual NVSS source redshifts are unknown, for our purposes
we need only the overall redshift distribution $dN/dz$ for the NVSS.
We adopt the favored model of \citet{dunlop/peacock:1990}
(model 1, MEAN-$z$ data), which was found by \citet{boughn/crittenden:2002}
to best reproduce the NVSS auto-correlation function.
Like most models, it divides the sources by
spectral index $\alpha$ (flux $S\propto\nu^{-\alpha}$) into two populations,
flat-spectrum ($\alpha\approx0$) and steep-spectrum ($\alpha\approx 0.8$)
sources.
The model is plotted in \reffigt{fig:dndz}.
We limit the model to $0.01<z<5$; the lower limit corresponds to a distance of
$\approx 42\,\textrm{Mpc}$.
The small peak at $z\approx 0.05$ is spurious
(due to a breakdown in the DP90 fitting function),
but has only a minor effect on the integrated predictions.

Given a cosmology, we can determine the radio bias $b_r$ from the amplitude of
the NVSS auto-correlation function (ACF), by comparing the ACF
to the unbiased prediction
\begin{equation}
C^{NN}(\theta)
= \langle \delta N(\nhat) \delta N(\nhat') \rangle
= \sum{\frac{2l+1}{4\pi} [b^N_l]^2 C^{NN}_l P_l(\nhat\cdot\nhat')}
\end{equation}
where $[b^N_l]^2$ is the pixel window function.
Substituting \refeqnt{eqn:dN-NVSS} into this expression and Fourier
transforming,
\begin{equation}
C^{NN}_l = 4\pi\int{\frac{dk}{k}\, \Delta^2_\delta(k) \left[f^N_l(k)\right]^2}
\end{equation}
where $\Delta^2_\delta(k) = k^3P_\delta(k)/2\pi^2$ is the logarithmic matter
power spectrum and $P_\delta(k) = \langle|\delta(k)|^2\rangle$.
We use the {\WMAP} normalization
$\Delta^2_{\cal R}(k_0) = 2.95\times10^{-9}A$ where $k_0=0.05\,{\rm Mpc}^{-1}$
\citep{verde/etal:2003},
giving us $\delta_H=6.1\times10^{-5}$ for our fiducial model.
The filter function is given by
\begin{equation}
f^N_l(k) = b_r \int{dz\, \frac{dN}{dz} D(z) j_l(k\eta)}.
\end{equation}
where $j_l(x)$ is the spherical Bessel function.
Note that the bias we measure is complicated by the uncertainty in
$dN/dz$; errors in the normalization are absorbed into the bias.

To calculate the observed NVSS auto-correlation function (ACF), we
made a HEALPix \citep{gorski/hivon/wandelt:1998}
resolution-5 map of $\delta N(\nhat)$,
which has $12\,288\ 1.8\arcdeg$~square pixels\footnote{For
more information on HEALPix visit 
\url{http://www.eso.org/science/healpix/}.
The \textit{resolution} of a HEALPix map indicates its pixel count.
A resolution-$r$ map has $12 N_{\rm side}^2$ pixels,
where $N_{\rm side}=2^r$. A resolution-$(r+1)$ map is created by dividing
each resolution-$r$ pixel into four subpixels.}.
As a precaution, we removed the $3\times10^5$ sources from the catalog
which were resolved.
The mean source count per pixel is 147.9,
leading to a Poisson uncertainty of $\approx 8\%$.
The ACF is estimated as
\begin{equation}
\hat{C}^{NN}(\theta_k) = \sum{N_i N_j w^N_i w^N_j}/\sum{w^N_i w^N_j}
\end{equation}
where $N_i$ is the number of sources in pixel $i$ and the sum
is over all pixel pairs separated by
$\theta_k-\Delta\theta/2<\theta<\theta_k+\Delta\theta/2$.
The bin width $\Delta\theta$ is $2\arcdeg$.
We mask out pixels at low Galactic latitude ($|b|<10\arcdeg$)
and those unobserved by the survey ($\delta<-37\arcdeg$);
the weights $w^N_i$ are determined by the mask.
The NVSS ACF is shown is \reffigt{fig:dndz}.
The $\theta=0\arcdeg$ bin is corrected for Poisson noise by subtracting
the mean number of sources per pixel.
For the fiducial model parameters (which are used to calculate $D(z)$
and $\eta(z)$), the derived bias is 1.7.
This is somewhat higher than the value of 1.6 found by
\citet{boughn/crittenden:2002}.
However, they assumed a scale-invariant spectrum (i.e., $n_s=1$),
and changing $n_s$ by $\pm 0.03$ changes the bias by $\mp 0.05$.

\begin{figure}
\plotone{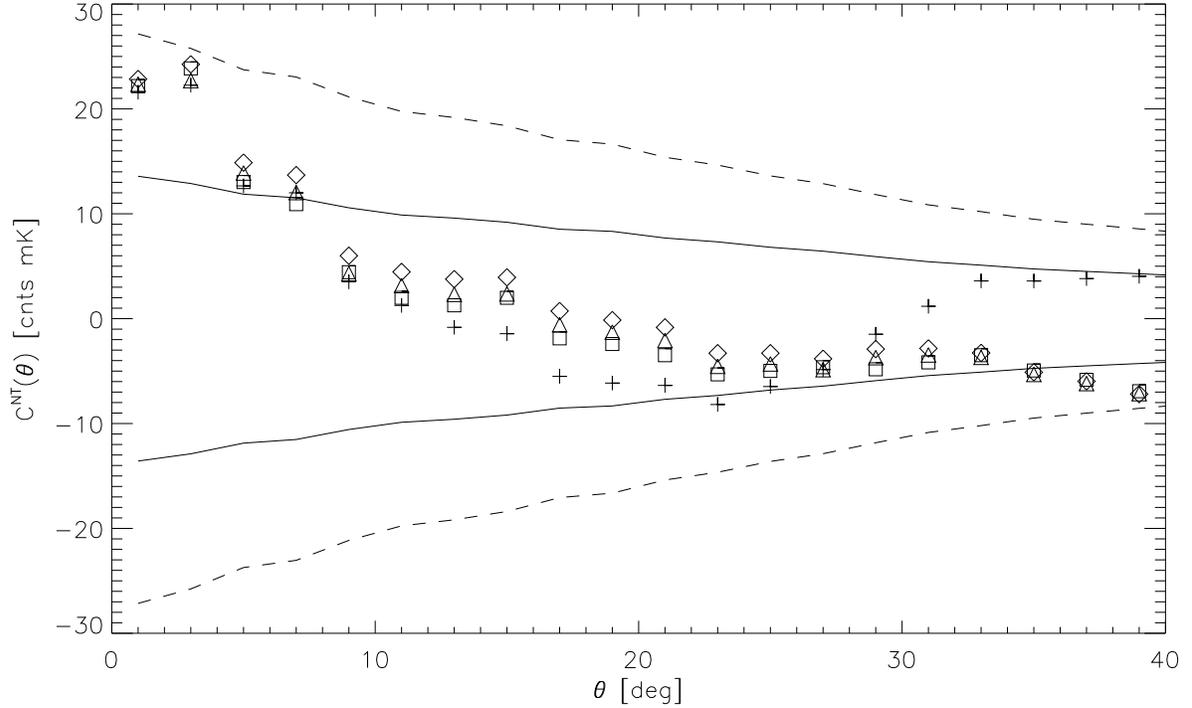}
\caption{\label{fig:wmap-nvss-ccf}
The \WMAP--NVSS cross-correlation function (CCF).
The CCF is insensitive to the details of the declination
correction. Two simple methods are compared; both broke the sources
into $\sin(\delta)$ strips of width $0.1$.
The first (diamonds) subtracted the mean from each strip.
The second (triangles) scaled each strip by the ratio of the global mean to the
strip mean.
The cross points are uncorrected,
showing the correction is only important for $\theta\ga 25$.
We used the {\WMAP} internal linear combination (ILC) CMB map;
substituting the map of \citet{tegmark/deoliveira-costa/hamilton:2003}
instead produces the same results (square points).
The solid and dashed lines are derived from the diagonal elements of the
correlation matrix due to accidental alignments;
they would be the $1\sigma$ and $2\sigma$ contours in the absence of
off-diagonal correlations.
The points, however, are highly correlated as shown in
Table~\ref{tbl:sim_corr_matrix}.
}
\end{figure}

As noticed by \citet{boughn/crittenden:2002}, the NVSS catalog
mean source density varies with declination, introducing a spurious signal
into the auto-correlation function. They corrected for this by
adding and subtracting random sources from the map until the structure
was removed.
We considered two simple corrections;
both broke the sources into $\sin(\delta)$ strips of width $0.1$.
The first method subtracted the mean from each strip;
the second scaled each strip by the ratio of the global mean to the
strip mean.
Since the corrections are small, both produced similar results
\reffigp{fig:wmap-nvss-ccf}.

\section{\WMAP--NVSS Cross-Correlation}

Combining \refeqnt{eqn:dT-ISW} and \refeqnt{eqn:dN-NVSS} we can calculate
the expected cross-correlation spectrum between the NVSS catalog
and the CMB:
\begin{equation}
C^{NT}_l = \langle a^N_{lm} a^{T*}_{lm} \rangle
	= 4\pi\int{\frac{dk}{k}\,\Delta^2_\delta(k) f^{N}_l(k) f^{T}_l(k)}
\label{eqn:clnt}
\end{equation}
where $f^{N}_l$ and $f^{T}_l$ are the NVSS and ISW filter functions.
The ISW filter function is derived analogously to the NVSS filter
function.
The local gravitational potential is related to the matter distribution
via the Poisson equation $\nabla^2\Phi = 4\pi G a^2 \rho_m \delta_m$,
where the gradient is taken with respect to comoving coordinates.
Fourier transforming, we have
\begin{equation}
\Phi(k,\eta) = -\frac{3}{2}\Omega_m (H_0/k)^2 g(\eta) \delta(k)
\end{equation}
where $H_0$ is the Hubble constant, $\Omega_m H^2_0=8\pi G\rho^0_m/3$,
and $g(\eta)\equiv D(\eta)/a(\eta)$ is the linear growth suppression factor.
Thus
\begin{equation}
f^T_l(k) = 3\Omega_m (H_0/k)^2 
	\int{d\eta\, \frac{dg}{d\eta} j_l(k\eta)}.
\label{eqn:filter-ISW}
\end{equation}
We use the fitting function for $g(\eta)$ provided by
\citet{carroll/press/turner:1992}.
In a flat $\Omega_m=1$ universe, $g(\eta)$ is constant, and thus there is
no ISW effect and hence no correlation between the CMB and the local matter
distribution.
In $\Lambda$CDM universes, $D(\eta)$ approaches a constant during
$\Lambda$-domination, leading to a decay of $g(\eta)$ as time increases.

We computed the \WMAP--NVSS cross-correlation function (CCF) as
\begin{equation}
\hat{C}^{NT}(\theta_k) = \sum{N_i T_j w^N_i w^T_j}/\sum{w^N_i w^T_j}
\label{eqn:measured-ccf}
\end{equation}
where $T_i$ is the CMB map, $N_j$ the NVSS map (described in the previous
section), and the sums are over all pixel pairs separated by
$\theta_k-\Delta\theta/2<\theta<\theta_k+\Delta\theta/2$.
The bin width $\Delta\theta$ is $2\arcdeg$.
As before, we work at HEALPix resolution-5.
Since we are working at large scales where the detector noise is
negligible, we use the {\WMAP} internal linear combination (ILC) map
for the CMB \citep{bennett/etal:2003b}.
We limit residual foreground contamination by masking the map
with the {\WMAP} Kp0 galaxy mask and the {\WMAP} source mask
\citep{bennett/etal:2003c}.
The CMB weight $w^T_i$ is the number of unmasked resolution-9 subpixels of the
resolution-5 pixel~$i$.
The CCF is plotted in \reffigt{fig:wmap-nvss-ccf}.
The CCF is insensitive to the form of the NVSS
declination correction. 

Assessing the significance of the CCF is complicated by the high
degree of correlation between points.
Accidental alignments between the NVSS map and the CMB fluctuations at the
decoupling surface (which are uncorrelated with those generated by the
late-ISW) can produce spurious correlations.
We quantified this uncertainty with Monte Carlo simulations,
creating 500 realizations of the CMB sky drawn from the $C^{TT}_l$
power spectrum for our fiducial model
with our estimate of the NVSS-correlated late-ISW contribution subtracted.
The covariance matrix $\Sigma$ was calculated from the resulting CCFs,
keeping the NVSS map fixed.
\reftbl{tbl:sim_corr_matrix} shows the $0\arcdeg<\theta<20\arcdeg$ submatrix of
$\Sigma$, showing that the CCF points are highly correlated.
We define $\chi^2 = \delta C^T\Sigma^{-1}\delta C$
where $\delta C=\hat{C}^{NT}-C^{NT}$
is the difference between observed \refeqnp{eqn:measured-ccf}
and model \refeqnp{eqn:clnt} correlation functions,
and we limit $\theta<20\arcdeg$.
For the null model of no correlation ($C^{NT}=0$), $\chi^2_0=\nullchisq$.
Since there are 10 degrees of freedom this a $1.8\sigma$ deviation.

\begin{figure}
\plotone{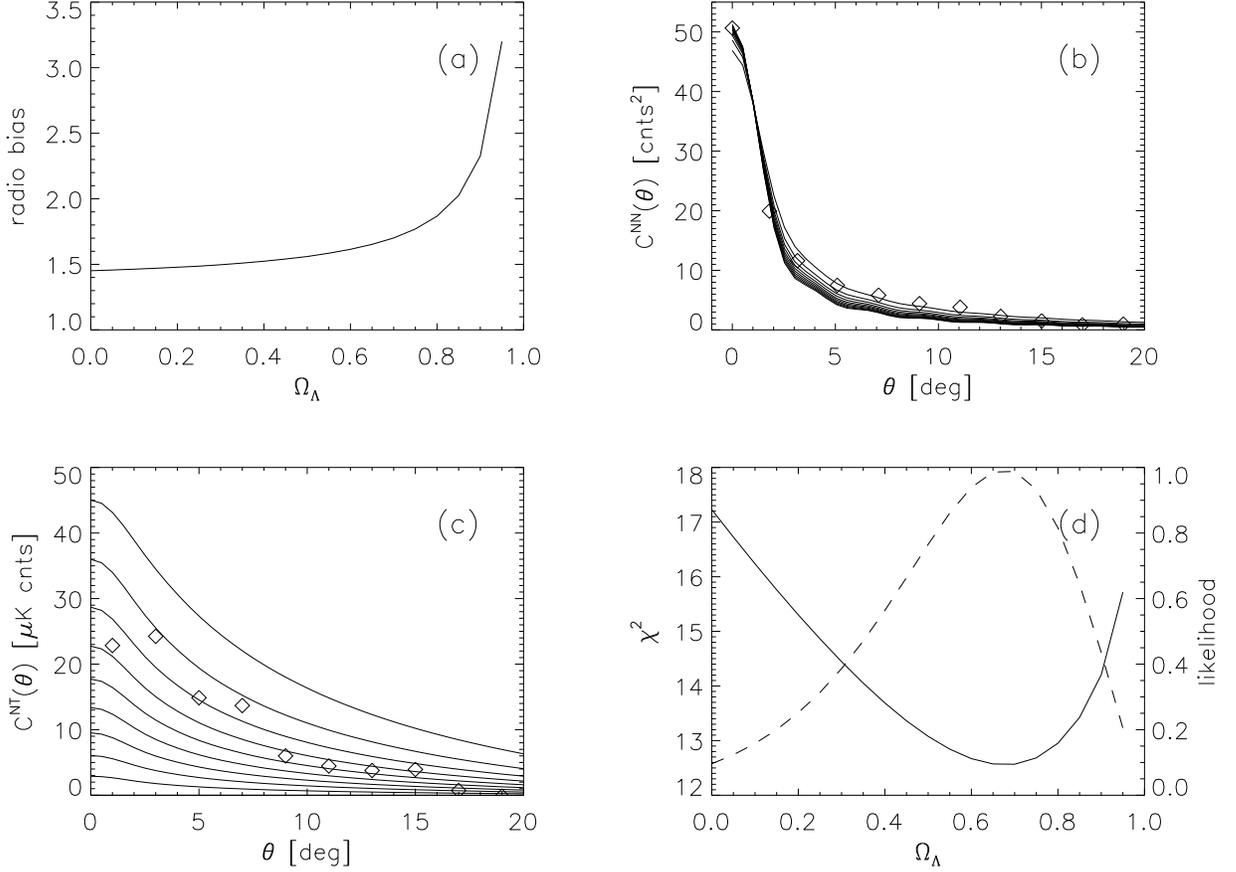}
\caption{\label{fig:lambda}
Effect of varying $\Omega_\Lambda$ on the cross-correlation function.
In all panels we assume a flat universe with fixed $\omega_b$, and
trade off between $\Omega_\Lambda$ and $h$ by keeping the combination
$\Omega_m h^{3.4}$ constant; when $\Omega_\Lambda=0$, $h=0.48$.
Panel (a) shows the inferred radio bias as a function of $\Omega_\Lambda$.
Panel (b) shows the bias-corrected NVSS auto-correlation function (ACF)
compared with the measured ACF.
Panel (c) shows the predicted cross-correlation function (CCF)
for a range of values of $\Omega_\Lambda$,
compared with the measured CCF (diamonds).
The amplitude of the predicted CCF is proportional to $\Omega_\Lambda$,
which is stepped in increments of 0.1 from 0.0 to 0.9.
Panel (d) shows the $\chi^2$ of the model CCF as a function of
$\Omega_\Lambda$.
The $\chi^2$ was computed using the first 10 points of the CCF
($0\arcdeg<\theta<20\arcdeg$) and Table~\ref{tbl:sim_corr_matrix}.
The minimum $\chi^2_{\rm min}$ is $\minchisq$ at $\Omega_\Lambda=\bestomegal$;
at $\Omega_\Lambda=0$, $\chi^2_0=\nullchisq$.
The dashed line is the likelihood $\propto \exp(-\chi^2/2)$.
The $1\sigma$ limits are $0.42<\Omega_\Lambda<0.86$.
}
\end{figure}

How does the measured CCF constrain $\Omega_\Lambda$?
Rather than exploring the full parameter space, 
we assume a flat universe with fixed $\omega_b$ and explore the locus
of values of $\Omega_m$ and $h$ consistent with
the measured location of the first acoustic peak of the
CMB TT anisotropy power spectrum.
The first peak position is set by the angular scale 
of the sound horizon at decoupling, $\theta_A$, which \citet{page/etal:2003c}
found to be $\theta_A=0.6\arcdeg$.
\citet{percival/etal:2002} showed that
$\theta_A \approx 0.85\arcdeg \Omega_m^{0.14} h^{0.48}$;
this is the horizon angle degeneracy.
The normalization $A$ is varied to fix the amplitude of the first peak;
from a fit to CMBFAST \citep{seljak/zaldarriaga:1996} spectra
we found $A\propto\Omega_m^{0.248}$ along the horizon degeneracy.
The results are shown in \reffigt{fig:lambda}.
The difference in $\chi^2$ between models with $\Omega_\Lambda=0$
(i.e., no correlation) and $\Omega_\Lambda=\bestomegal$
(the minimum, with $\chi^2_{\rm min}=\minchisq$) is
$\Delta\chi^2=\deltachisq$.
Since we are varying a single parameter, the significance is
$\sqrt{\Delta\chi^2}$;
thus $\Omega_\Lambda>0$ is preferred at the $2.2\sigma$ level.

The {\WMAP} team imposed a prior on the Hubble constant, $h>0.5$, in
determining the cosmological parameters \citep{spergel/etal:2003}.
While lower values of the Hubble constant would contradict a host of
other experiments, especially the Hubble Key Project \citep{freedman/etal:2001},
models with very low $h$ and $\Omega_m\approx1.3$
are marginally consistent ($\Delta\chi^2=4.9$)
with the {\WMAP} TT and TE angular power spectra.
Since these universes are closed and matter dominated
the growth factor $D(a)$ grows faster than $a$, 
and $g(a)$ is better termed the linear growth \textit{enhancement} factor.
Thus we would expect to observe an anti-correlation between the NVSS
and CMB maps, since $dg/d\eta$ in \refeqnt{eqn:filter-ISW} changes sign
(see \reffigt{fig:phi}).
For $\Omega_m=1.28$ and $h=0.33$, $\chi^2=24.2$ (with a bias of $2.7$).
Based on the cross-correlation analysis alone,  this model is
disfavored by more than $3\sigma$.

\section{Discussion}

The recent acceleration of the universe due to dark energy
should correlate large-scale CMB anisotropies with fluctuations in
the local matter density through the late-time integrated Sachs-Wolfe
effect.
We have correlated the NVSS radio source catalog with the CMB
anisotropies observed by the {\WMAP} satellite, and
find that $\Omega_\Lambda>0$ is preferred at the 95\% confidence level
($\Delta\chi^2=\deltachisq$), considering statistical errors only.
The statistical uncertainty is due to accidental alignments with the background
primary anisotropies generated at decoupling.
The likelihood peaks at $\Omega_\Lambda=\bestomegal$, consistent with the value
derived from the CMB angular power spectrum.

The correlation between the NVSS source count and {\WMAP} CMB maps appears
robust.
We interpret it as arising from the late-ISW,
but other effects could correlate the two maps.
For instance, obscuration by dust clouds
tends to reduce the number of sources observed in their direction.
We cross-correlated the NVSS map with the $E(B-V)$ extinction map of
\citet{schlegel/finkbeiner/davis:1998} and see evidence for
a small negative correlation at separations $\theta<20\arcdeg$.
However, since the extinction map is positively correlated with the CMB
map due to dust emission, this effect has the wrong sign to mimic
the late-ISW.
The extinction correction is estimated as
$r_E(\theta)=\langle EN\rangle\langle ET\rangle/\langle EE\rangle$,
where $N,T,E$ are the NVSS, CMB, and extinction maps and $\langle XY\rangle$
denotes the correlation between maps $X$ and $Y$ evaluated at
separation $\theta$.
We find a value of $r_E(0)\approx -4\,\rm \mu K\,cnts$
(compared to $22\,\rm \mu K\,cnts$ for the CCF)
and the correction is negligible for $\theta>15\arcdeg$, 
except for a few glitches when $\langle EE\rangle$ crosses zero.
Subtracting $r_E$ from the CCF, we find the preferred value of
$\Omega_\Lambda$ increases to $0.76$;
at the minimum $\chi_{\rm min}^2=14.8$, and $\Delta\chi^2=6.1$.

A potentially more serious concern is that the correlation is due to microwave
emission by the sources themselves.
However, if this were the case then the CCF should have a similar
angular profile as the auto-correlation function (ACF).
Yet while the ACF falls steeply with increasing separation
[$\hat{C}^{NN}(0\arcdeg)/\hat{C}^{NN}(3\arcdeg)\sim 5$], the CCF does
not [$\hat{C}^{NT}(0\arcdeg)/\hat{C}^{NT}(3\arcdeg)\sim 1$].
The lack of an enhanced signal in the zero lag CCF bin thus argues against
any significant microwave emission from the NVSS radio sources.

\acknowledgements

We are indebted to Steve Boughn for significant and helpful discussions
throughout the preparation of this paper. We also thank Hiranya Peiris
and Licia Verde for useful comments.
The {\WMAP} mission is made possible by the support of the Office of Space 
Sciences at NASA Headquarters and by the hard and capable work of scores of 
scientists, engineers, technicians, machinists, data analysts, 
budget analysts, managers, administrative staff, and reviewers. 


\clearpage

\begin{deluxetable}{ccccccccccc}
\tabletypesize{\small}
\tablecaption{\label{tbl:sim_corr_matrix}
\WMAP--NVSS CCF Correlation Matrix}
\tablehead{
\colhead{Bin}&
\colhead{1\arcdeg}&
\colhead{3\arcdeg}&
\colhead{5\arcdeg}&
\colhead{7\arcdeg}&
\colhead{9\arcdeg}&
\colhead{11\arcdeg}&
\colhead{13\arcdeg}&
\colhead{15\arcdeg}&
\colhead{17\arcdeg}&
\colhead{19\arcdeg}
}
\startdata
       1\arcdeg &
183&
159&
141&
131&
115&
98.7&
87&
73.7&
58.7&
49.7\\
       3\arcdeg &
&
161&
140&
136&
119&
104&
94.3&
82.8&
67.6&
59.6\\
       5\arcdeg &
&
&
135&
128&
113&
97.8&
86.4&
74.8&
60.4&
52.5\\
       7\arcdeg &
&
&
&
135&
120&
106&
96.7&
86.4&
71.9&
64.2\\
       9\arcdeg &
&
&
&
&
113&
102&
92.6&
83.7&
70.3&
63.4\\
      11\arcdeg &
&
&
&
&
&
95.9&
89.2&
81.6&
69.6&
63.5\\
      13\arcdeg &
&
&
&
&
&
&
88.3&
82.7&
71.7&
66.3\\
      15\arcdeg &
&
&
&
&
&
&
&
81.7&
72.3&
67.6\\
      17\arcdeg &
&
&
&
&
&
&
&
&
67.5&
64.2\\
      19\arcdeg &
&
&
&
&
&
&
&
&
&
64.6\\
\enddata

\tablecomments{
The correlation matrix $\Sigma$ of the \WMAP--NVSS cross-correlation function
due to accidental alignments with the anisotropies produced at
decoupling. Units are $(\rm\mu K\,cnts)^2$.
}
\end{deluxetable}

\clearpage

\end{document}